\documentclass{article}

\usepackage{amsmath}
\usepackage{amssymb}
\usepackage{graphicx}
\usepackage{subfigure}
\usepackage{cite}

\begin{document}

\begin{titlepage}

\begin{flushright}
{\tt RIKEN-TH-54} \\
{\tt hep-th/0510105}
\end{flushright}

\vfill

\begin{center}
{\Large\bf
Improved Taylor Expansion method in the Ising model }

\vfill

{{\sc Aoyama} Tatsumi
}, 
{{\sc Matsuo} Toshihiro
}, 
and
{{\sc Shibusa} Yuuichirou
}

\vfill

{\it 
Theoretical Physics Laboratory,\\
The Institute of Physical and Chemical Research (RIKEN),\\
Wako, 351-0198, Japan
}
\end{center}

\vfill

\begin{abstract}
We apply an improved Taylor expansion method, which is a 
variational scheme to the Ising model in two dimensions. 
This method enables us to evaluate the free energy and magnetization 
in strong coupling regions from the weak coupling expansion, 
even in the case of a phase transition. 
We determine the approximate value of the transition point using this scheme. 
In the presence of an external magnetic field, we find both stable 
and metastable physical states. 
\end{abstract}

\vfill

\end{titlepage}

\paragraph{Introduction.}

More often than not, we encounter situations in which it is 
difficult to
evaluate physical quantities by means of standard perturbation 
methods because, for example, the theory under consideration does 
not have any small parameters in which we can expand. 
As non-perturbative methods, variational schemes 
have been applied to such circumstances with great success. 
In such methods one or more auxiliary parameters are introduced 
into the model. 
Optimized perturbation theory \cite{Yukalov,Stevenson:1981vj} 
is a systematic improvement of variational methods, 
formulated on the basis of the ``principle of minimal sensitivity''. 
\cite{Stevenson:1981vj}
In zero and one dimensions, it has been proved that the optimized series 
converges. \cite{Guida:1994zv}\ 
This method has been applied to, among others, matrix models of 
superstring theory 
and their simplified toy models. 
\cite{%
Kabat:1999hp,%
Oda:2000im,
Nishimura:2002va,Nishimura:2003gz,Nishimura:2004ts,%
Nishimura:2001sx,Kawai:2002jk,Kawai:2002ub}
It was first reported in %
Ref.~\cite{Kawai:2002jk} 
that minimal sensitivity is realized in auxiliary parameter space 
as a {\em plateau} 
(i.e. a region in which physical quantities are stable), 
and that the appearance of a plateau can be regarded as a signal 
reflecting whether or not the method works. 
It has also been argued that the optimized perturbation theory can 
be explicitly formulated as an improved Taylor series, 
which is obtained in standard perturbation theory. 

In this article, we attempt to improve our understanding of this 
method through application to the Ising model in 
two dimensions.\footnote{%
In Ref.~\cite{Shino} there appears a treatment of the 
application of the improved Taylor expansion method to the Ising model. 
Our arguments here extend and clarify the work given there. 
}
It is generally believed that 
physics in an ordered phase cannot be analyzed using 
a na\"{i}ve perturbation theory formulated in a disordered phase, 
because, in general, 
a perturbation theory yields an ill-defined 
expansion series when it is applied to a different phase than that 
in which it was constructed. 
We utilize a mathematical technique to reorganize the series 
resulting from a perturbation theory 
in a non-trivial manner which is known to yield 
a good approximation even 
in the case that the expansion is carried out about a point 
outside the radius of convergence.\cite{Kawai:2002jk} 
We show that our scheme actually enables us to derive 
information concerning the ordered phase from a perturbation theory 
formulated in the disordered phase, and vice versa. 

We do not attempt here to calculate the series up to as high an order as 
possible in order to obtain more accurate estimates in the 
approximation scheme we consider. 
Rather, our goal is to show that physical information 
can be extracted from relatively low orders of perturbation 
with this method. 
In actual problems for which the method is intended, such as QED and QCD, 
no more than first three or four terms can be computed in perturbation 
theory, and from these, we are required to extract physical information.%
\footnote{%
One attempt to apply this method to non-abelian Yang-Mills theory
can be found in Ref.~\cite{Kawamoto:2003kn}.
}

In the following, we first briefly describe how 
the optimized perturbation method can be understood as an improved Taylor 
expansion (ITE). 
Then we can apply it to the Ising model 
in two dimensions to determine whether the method works in a system 
that exhibits a phase transition. 
By applying the ITE to this model, 
we show that the ITE provides a phase 
which is invisible from another phase in the perturbation expansion. 
We also investigate within this model the extent to which 
we are able to understand the critical behavior from the ITE analysis.

\paragraph{Improved Taylor expansion.}

Suppose we have some quantity 
$F(\lambda, m)$ which is an implicit function of the parameters
$m$ and $\lambda$, 
and we would like to evaluate it
at some definite values of these parameters.
In order to treat $F(\lambda, m)$ using perturbation techniques, 
we regard one of the parameters, $\lambda$, as an expansion parameter. 
However, at this point we do not assume that $\lambda$ 
is necessarily small. Also, in general, 
the parameter $m$ might respect a set of parameters, 
but we will not consider such cases in this article. 
Then, one might perform a formal perturbative expansion 
with respect to $\lambda$ up to some finite order, say, $N$th 
order, which depends on the nature of the problem at hand 
(or our ability to carry out the necessary calculations). 
From the $N+1$ coefficients appearing in the expansion, 
we obtain a sequence of finite series $F_0, F_1, \cdots, F_N$ 
in the form 
\begin{equation}
	F_i(\lambda,m) = \sum_{n=0}^i \lambda^n f_n(m),
	\qquad
	0 \leq i \leq N \,.
\end{equation}
It is not necessarily the case that 
each such series is convergent, 
because the expansion parameter $\lambda$ might be large 
(outside the radius of convergence), 
or even for small $\lambda$, it could turn out to diverge 
at some order if it is an asymptotic series.
Our goal here is to improve the convergence of this series. 
For this purpose, we first introduce a formal expansion parameter $g$ and 
an auxiliary parameter $m_0$, 
and then make the substitutions 
\begin{align}
	\lambda &\to g\lambda \,, 
\nonumber \\
	m       &\to m_0+g(m-m_0) \,, 
\end{align}
in the arguments of each series $F_i(\lambda, m)$. 
Then, it is easily seen that 
\begin{align}
	\sum_{n=0}^i \lambda^n f_n(m)
	&=
	\sum_{n=0}^i g^n \lambda^n f_n(m_0+g(m-m_0)) \Big|_{g=1} 
\nonumber \\
	&=
	\sum_{n=0}^i \lambda^n
	\left(
		\sum_{k=0}^{i-n} g^{k+n}
		\frac{(m-m_0)^k }{k!} 
		\frac{\partial^k f_n(m_0)}{\partial m^k} 
		+{\cal O}(g^{i+1})
	\right)
	\Bigg|_{g=1} \,,
\label{eq:untruncated}
\end{align}
where the notation $\bigl|_{g=1}$ means that we substituted $g=1$ at the end.
It might not be obvious from the above expression that 
the last expression of (\ref{eq:untruncated}) is independent 
of the parameter $m_0$ as a whole, but in fact they are. 
However, if we drop the ${\cal O}(g^{i+1})$ terms, 
the coefficient functions become dependent on $m_0$. 
Following these steps, we obtain a new sequence of series 
which we call the ``improved functions'':
\begin{equation}
	F_i^{imp}(\lambda,m; m_0) =
	\sum_{n=0}^i \lambda^n
	\left(
		\sum_{k=0}^{i-n} 
		\frac{(m-m_0)^k }{k!} 
		\frac{\partial^k f_n(m_0)}{\partial m^k} 
	\right)
	\,.
	\quad 
	(i=0, \cdots, N)
\end{equation}
According to the principle of minimal sensitivity, 
\cite{Stevenson:1981vj}\ 
the improved functions best approximate the exact value 
when they are evaluated in regions where they are 
least sensitive to the auxiliary parameter $m_0$. 
We refer to the regions in which the minimal sensitivity 
condition is realized as {\em plateaux}. \cite{Kawai:2002jk}

It is a rather difficult problem to identify a plateau, 
especially when there are many auxiliary parameters.\footnote{%
In 
Refs.~\cite{Kawai:2002jk,Kawai:2002ub,Aoyama-Kawai} 
the accumulation of extremal points 
in the parameter space is used as a criterion for the identification 
of a plateau. 
An efficient method to detect a plateau in a multi-dimensional parameter 
space is presented in Ref.~\cite{Nishimura:2002va}, 
where the histogram method is proposed.
Further discussion will be given in a forthcoming paper.
\cite{Aoyama-Shibusa}\ 
}
Indeed, we have no quantitative nor mathematically rigorous 
definition of a plateau. 
However, in the case of a single auxiliary parameter, 
one may plot graphs of the improved series as functions of the parameters. 
Then, examining these graphs, we are able, 
at least qualitatively, to determine whether or not a plateau exists.

\paragraph{Ising model.}

The Ising model \cite{{Domb-Green}} is one of the simplest models that 
exhibit phase transitions. 
Here, we test the ITE using the Ising model 
on a two-dimensional square lattice 
with a uniform magnetic field $h$. 
The total number of sites is denoted by $V$, which defines 
the volume of the system.
The Hamiltonian is given by 
\begin{equation}
	H = -J \sum_{\langle ij \rangle} \sigma_i \sigma_j 
	- h \sum_i \sigma_i \,,
\end{equation}
where $\sigma_i$ located at the site labeled by the index $i$ takes 
one of the values $\pm1$, and there are interactions 
only between nearest neighbors, with the strength $J\ (>0)$.  
Pairs of neighboring sites are denoted as $\langle ij \rangle$. 
The partition function is defined as 
\begin{equation}
	Z(J,h) = \sum_{\{\sigma\}} e^{-H} \,,
\end{equation}
where we normalize the coupling and external field strength 
in units of $k_B T$. 
The sum is taken over all possible configurations. 

The exact free energy of the two-dimensional Ising model 
was obtained by Onsager: \cite{Onsager:1943jn}
\begin{align}
	f_{\rm exact}(J)
	=&
	\ln\left(\frac{1-k^2}{2}\right) 
\nonumber \\
	& -\frac{1}{8\pi^2}
	\int_0^{2\pi}\!\!dx\int_0^{2\pi}\!\!dy\, 
	\ln\left[(1+k^2)^2-2k(1-k^2)(\cos x +\cos y)\right] \,,
\label{eq:exactf}
\end{align}
where $k=\tanh(J)$.
Also, the exact expression of the magnetization \cite{Onsager-Yang}
is 
\begin{equation}
	m(J)=\left[1-\sinh^{-4}(2J)\right]^{1/8} .
\label{eq:exactmag}
\end{equation}

In the following, we first apply the ITE to the weak coupling
expansion and then to the strong external magnetic field expansion.

\paragraph{Weak coupling expansion.}

In this section, 
we employ the weak coupling expansion\footnote{ 
This series can be understood as the high temperature expansion
in the case of zero external magnetic field, $h=0$.}
and construct the following series: 
\begin{equation}
	Z(J,h) \equiv \sum_{n=0}^\infty J^n z_n(h) \,.
\end{equation}
Calculating $z_n(h)$ up to fourth order in $J$ 
we have the following: 
\begin{subequations}
\begin{align}
	z_0(h)
	&= 
	(2\cosh(h))^V , \\
	z_1(h)
	&= 
	z_0 2V\tanh^2(h) , \\
	z_2(h)
	&= 
	\frac{z_0}{2!} 2V
	\left[
		(2V-7)\tanh^4(h)+6\tanh^2(h)+1
	\right] , \\
	z_3(h)
	&= 
	\frac{z_0}{3!} 2V
	\left[
		(4V^2-42V+116)\tanh^6(h)+(36V-168)\tanh^4(h)
\right. \nonumber \\ & \left.
		+(6V+52)\tanh^2(h)
	\right] , \\
	z_4(h)
	&= 
	\frac{z_0}{4!} 2V
	\left[
		(8V^3-168V^2+1222V-3126)\tanh^8(h)+(144V^2-1848V
\right. \nonumber \\ & \left.
		+6312)\tanh^6(h)+(24V^2+548V-3748)\tanh^4(h)+(72V+552)\tanh^2(h)
\right. \nonumber \\ & \left.
		+(6V+10)
	\right].
\end{align}
\end{subequations}
Note here we are ignoring the effect of the boundaries, 
which is irrelevant in the large volume (thermodynamic) limit.

The free energy density at $j$th order, $f_j(J, h)$, is obtained as 
a series expansion up to order $j$ of the logarithm of the partition function:
\begin{equation}
	f_j(J, h)=-\frac{1}{V}\ln\left(\sum_n J^n z_n(h)\right)\Bigg|_j \,.
\end{equation}
Here, the notation $\Bigl|_j$ indicates that all terms of order $j+1$ or 
higher in $J$ are omitted. 
Because we have calculated the partition function up to fourth order in $J$, 
we can easily obtain the fourth-order free energy density as 
a function of $J$ and $h$:
\begin{equation}
\begin{aligned}
	f_4(J, h)
	&= 
	-\ln(2\cosh(h))-2\tanh^2(h) J
	-\left\{1+6\tanh^2(h)-7\tanh^4(h)\right\}J^2 \\
	&
	-\left\{\frac{52}{3}\tanh^2(h)-56\tanh^4(h)+\frac{116}{3}\tanh^6(h)\right\}J^3 \\
	&
	-\left\{\frac{5}{6}+46\tanh^2(h)-\frac{937}{3}\tanh^4(h)+526\tanh^6(h)-\frac{521}{2}\tanh^8(h)\right\}J^4 \,.
\end{aligned}
\label{free4}
\end{equation}
This function is independent of $V$, as should be the case. 

The magnetization density is obtained 
by differentiating the free energy density 
with respect to the external magnetic field $h$:
\begin{equation}
	m_j(J, h) = - \frac{\partial f_j(J, h)}{\partial h} \,.
\end{equation}
This leads to 
\begin{equation}
\begin{aligned}
	& m_4(J, h)
	= 
	\tanh(h)\biggr[	
	1+4\{1-\tanh(h)^2\}J+4\{3-10\tanh(h)^2+7\tanh(h)^4\}J^2\biggl. 
\\ & \ 
	+\left\{\frac{104}{3}-\frac{776}{3}\tanh(h)^2+456\tanh(h)^4-232\tanh(h)^6\right\}J^3 
\\ & \ 
	+\left\{92-\frac{4024}{3}\tanh(h)^2+\frac{13216}{3}\tanh(h)^4-5240\tanh(h)^6+2084\tanh(h)^8\right\}J^4\biggl] .
\end{aligned}
\label{mag4}
\end{equation}
In the following, we refer to the free energy density 
and magnetization density as 
simply the free energy and magnetization.

\paragraph{Applying the ITE.}

Having thus obtained the basic ingredients 
through perturbative calculations about $J=0$, 
we now apply the ITE to the resulting series 
in order to obtain improved information, 
especially at large $J$ (low temperature). 
For this purpose, 
we introduce a formal expansion parameter 
$g$ and an auxiliary parameter $h_0$, 
and then make the substitutions 
\begin{align}
	J & \to g J \,, 
\nonumber \\
	h & \to h_0 + g( h-h_0) 
\end{align}
in the expressions for the 
free energy, (\ref{free4}), and magnetization, (\ref{mag4}). 
After carrying out a Taylor expansion about $g=0$ up to the specified order 
and then setting $g=1$, 
we obtain improved functions. 
In the present case, we have calculated the free energy 
and the magnetization up to fourth order in $g$. 
We have sequences containing five improved functions of 
the free energy, $f^{imp}_i(J, h;h_0)$, and magnetization, 
$m^{imp}_i(J, h;h_0)$, 
which contain the fictitious parameter $h_0$. 
We list the explicit forms of these improved functions 
in Appendix \ref{sec:appendix-A}.

In order to investigate the spontaneous magnetization, 
we set the external magnetic field, $h$,  to zero in the improved functions. 
The results obtained using the ITE can then be compared 
with the exact values, (\ref{eq:exactf}) and (\ref{eq:exactmag}), 
to test the validity of the method. 

Before discussing the improved functions in detail, 
let us first comment on the behavior of the functions 
in the case $h_0 = 0$ and $h_0 \to \pm\infty$. 
For $h_0=0$, the improved free energy at $i$th order 
coincides with the original one 
at the same order as a function of $J$; e.g., we have 
\begin{align}
	f^{imp}_4(J,0;0) 
	&= -\ln 2-J^2-\frac{5}{6}J^4 
\nonumber \\
	&= f_4(J,0) \,.
\end{align}
For the asymptotic behavior in the limits $h_0 \to \pm\infty$, 
all of the improved free energies 
become close to $-2J$ (except $f_{0}^{imp}$ which diverges):
\begin{equation}
	f^{imp}_i(J,h; h_0 \to \pm \infty) = \pm h -2J 
	\,,
	\qquad 
	(i=1,2,3 \mbox{ and } 4)
\end{equation}
where we have restored $h$ for later convenience. 
This shows that the improved free energies are insensitive to 
the value of the parameter $h_0$ for large $|h_0|$, and thus, 
two flat regions are formed. 
However, we do not regard those asymptotic flat regions as plateaux, 
because their forms are independent of the order of perturbation; 
we consider a flat region to be a {\em plateau} only if its form 
changes as the order increases. 
This also implies that 
in order to determine whether a given region is a plateau, 
we have to consider the whole sequence 
of improved series, rather than only the functions of highest order. 
\begin{figure}
\begin{center}
\subfigure[][]{%
	\includegraphics[scale=0.6]{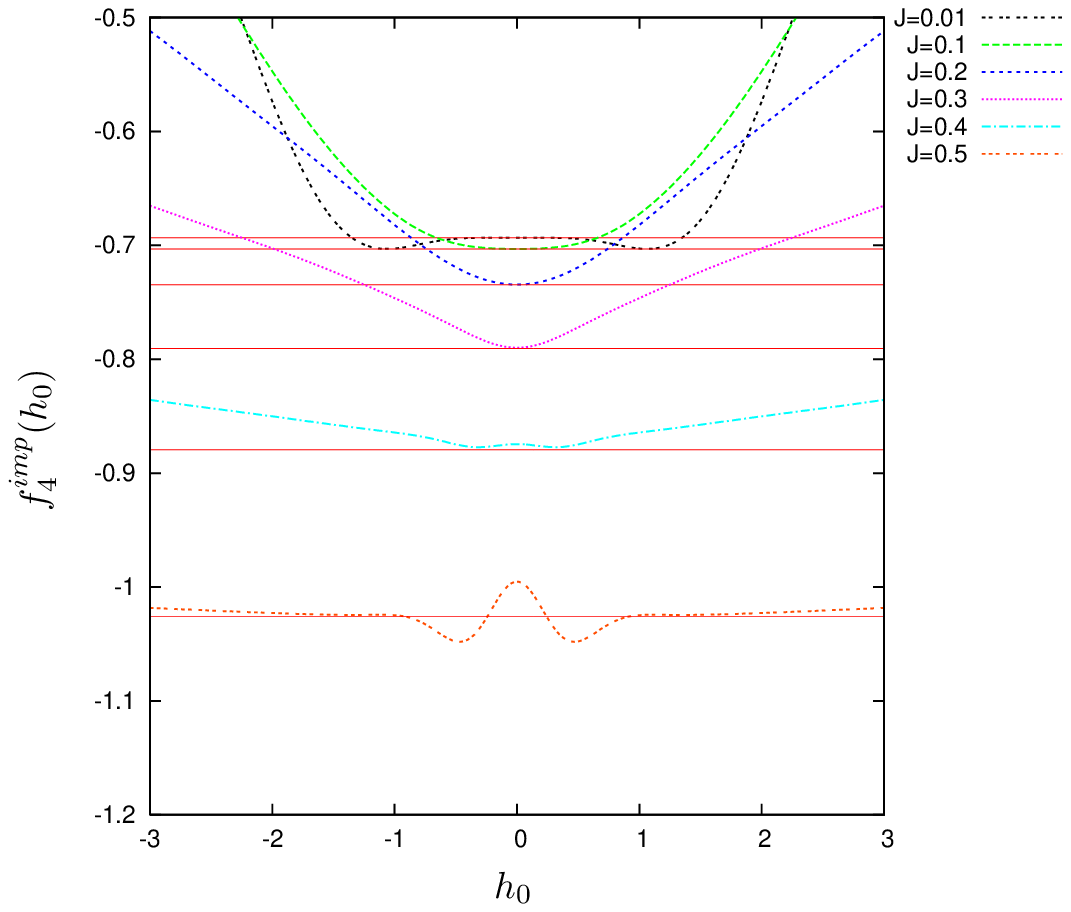}
\label{fig:01}
}
\hskip 4em
\subfigure[][]{%
	\includegraphics[scale=0.6]{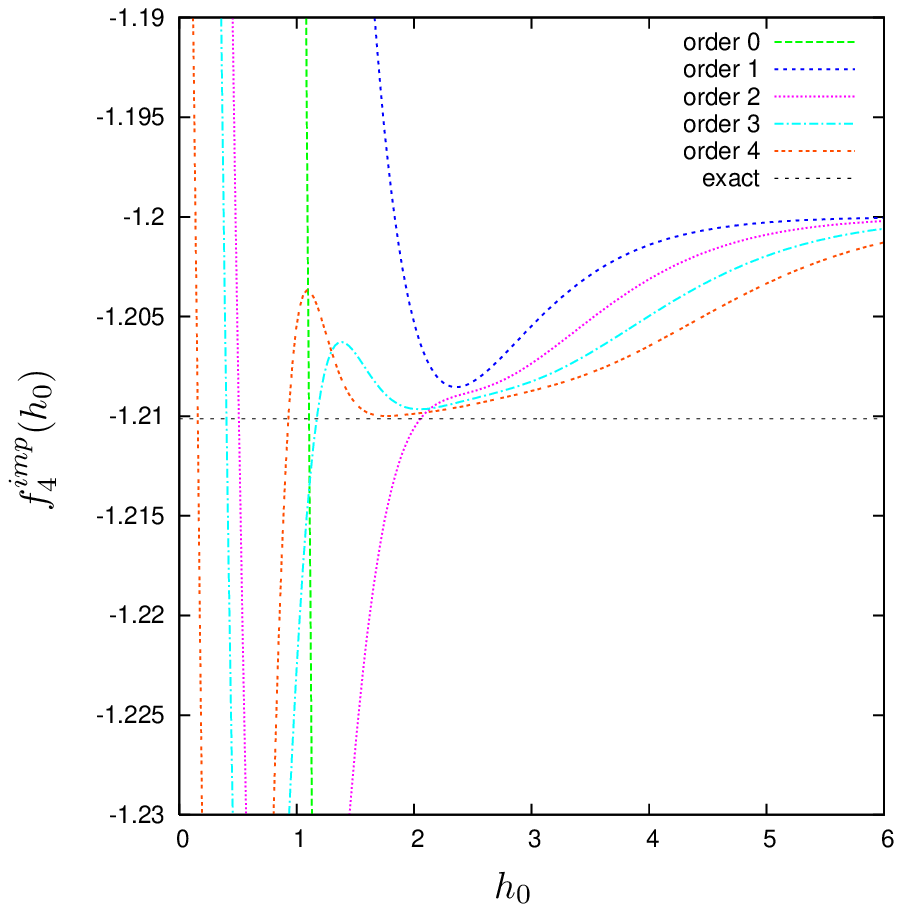}
\label{fig:02}
}
\end{center}
\caption{%
\subref{fig:01} %
Dependences of the fourth-order improved free energies, $f^{imp}_4(h_0)$, 
with respect to the auxiliary parameter $h_0$ 
for $J=0.01$, $0.1$, $0.2$, $0.3$, $0.4$, and $0.5$. 
\subref{fig:02} %
Dependences of the improved free energies $f^{imp}_i(h_0)$ 
for $J=0.6$ from zeroth order to fourth order. 
The horizontal lines represent the exact values of the free energies 
at the respective values of $J$.}
\end{figure}

With the above points in mind, 
let us study the behavior of the improved functions in detail. 
Figure \ref{fig:01} exhibits the fourth-order improved free energy 
$f_{4}^{imp}$ 
as a function of $h_0$ for $J=0.01$, $0.1$, $0.2$, $0.3$, $0.4$, and $0.5$. 
As we vary $J$, there is a clear evolution of 
the shape of the improved free energy. This reflects the fact that 
there are two phases, corresponding to the small $J$ (disordered) phase 
and the large $J$ (ordered) phase in the Ising model. 
At small values of $J$ (i.e. in the high temperature region), 
the set of improved free energies form a clear plateau 
around $h_0=0$, at which the values of the improved 
free energies are good approximations of the exact value. 
This is not surprising, 
because the original free energy was obtained 
through a perturbative expansion in small $J$, 
and it already provides a good approximation of the exact value. 
As we have seen above, at $h_0=0$, which is on the plateau, 
the fourth-order improved free energy has the same form as the original. 

Now, let us consider the case of relatively large $J$ 
(i.e. the low temperature region). 
In this case, it is seen that plateaux develop in the regions 
near $h_0 \sim \pm 2$ as 
the order of the improved free energy increases. 
Thus, we can evaluate the approximate value of the free energy 
on these plateaux, 
which is found to be close to the exact value expressed 
by the horizontal line [Fig.~\ref{fig:02}]. 
We would like to emphasize that we have obtained 
the relatively large $J$ (ordered phase) physics 
from the data of small $J$ (disordered phase) expansions. 

\begin{figure}
\begin{center}
	\includegraphics[scale=0.6]{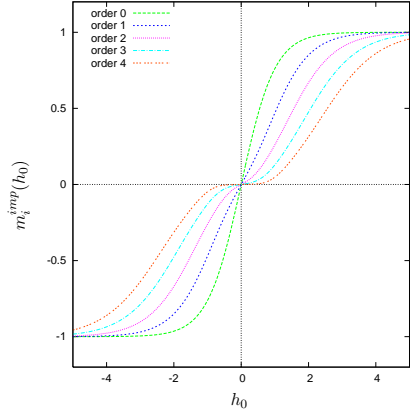}
\end{center}
\caption{%
Dependences of the improved magnetizations $m^{imp}_i(h_0)$ 
with respect to the auxiliary parameter $h_0$ at $J=0.1$.}
\label{fig:03}
\end{figure}
\begin{figure}
\begin{center}
\subfigure[][]{%
	\includegraphics[scale=0.6]{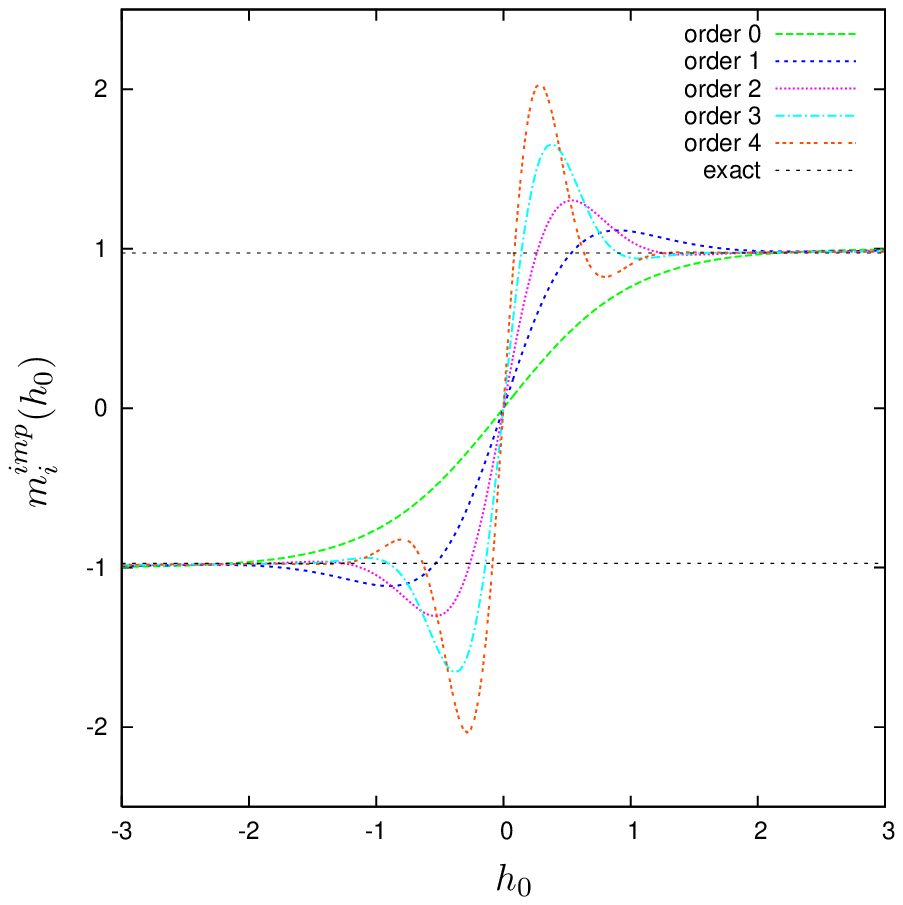}
\label{fig:04}
}
\hskip 2em
\subfigure[][]{%
	\includegraphics[scale=0.6]{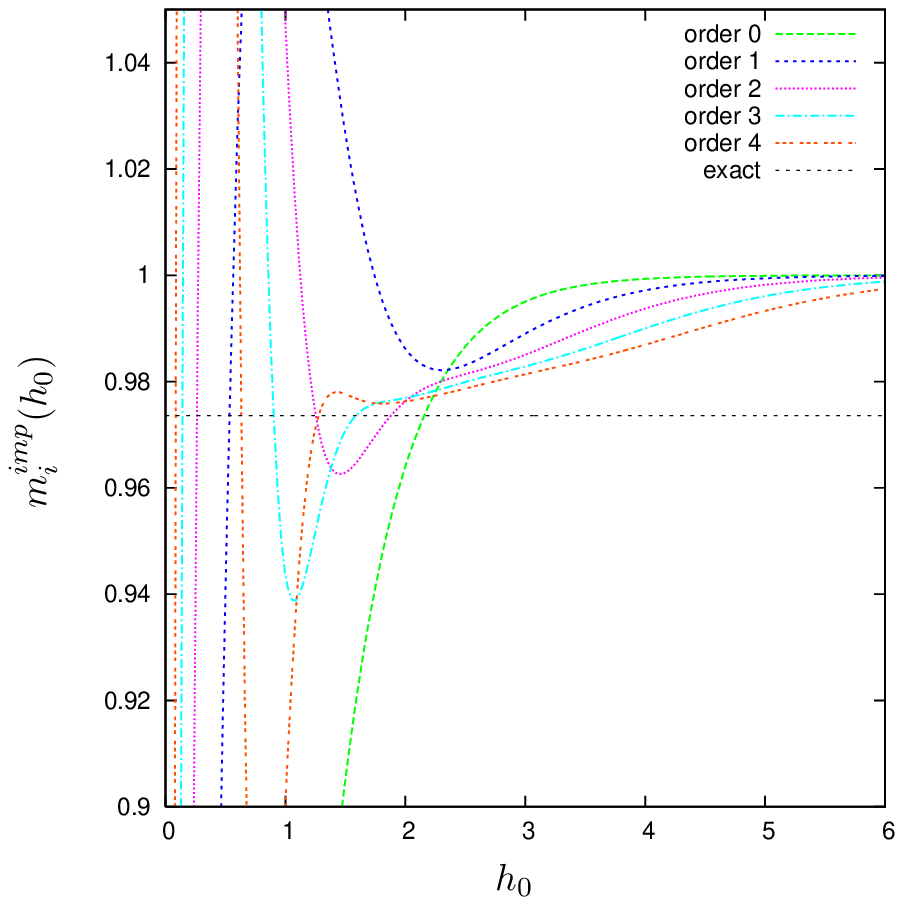}
\label{fig:05}
}
\end{center}
\caption{%
\subref{fig:04} %
Dependences of the improved magnetizations $m^{imp}_i(h_0)$ 
with respect to the auxiliary parameter $h_0$ for $J=0.6$. 
\subref{fig:05} %
An enlarged plot near the exact value, which is represented 
by the horizontal line.}
\label{fig:0405}
\end{figure}

Although the emergence of plateaux at non-zero $h_0$ 
in the improved free energies 
may signal the occurrence of spontaneous magnetization, 
we have to investigate the magnetization itself to determine 
the amount by which the system magnetizes spontaneously. 
In Figs.~\ref{fig:03} and \ref{fig:0405}, respectively, 
we plot the improved magnetizations 
from zeroth to fourth order for $J=0.1$ and $J=0.6$. 
In Fig.~\ref{fig:03}, it is seen that the slopes become gentle 
and a plateau emerges near the origin as the order increases. 
Contrastingly, Fig.~\ref{fig:04} shows that the slopes at the origin 
become steeper as the order increases, whereas two plateaux develop 
near $h_0 = \pm2$. 
This figure also provides an approximate value of 
the magnetization that is close to the exact value, 
represented by the horizontal line in Fig.~\ref{fig:05}. 

The two phases are characterized by spontaneous magnetization. 
If the plateau corresponds to the ground state realized at 
a particular value of $J$, 
the critical point can be recognized by the 
disappearance of the plateau formed at and near the origin. 
As mentioned above, the slope at the origin becomes steeper or 
gentler as we increase the order, depending on the value of $J$. 
Thus, at each order of approximation, the critical point is 
determined by the condition that the difference between the 
derivatives of the magnetizations at $n$th and $(n\!-\!1)$th orders 
be zero at the origin. 
Thus, we find\footnote{%
The critical temperature at first order ($J_c^{(1)}=0.25$) 
is the same as that calculated in the mean field approximation.}
\begin{subequations}  
\begin{alignat}{2}
	\left[\frac{\partial m^{imp}_1}{\partial h_0}-\frac{\partial m^{imp}_0}{\partial h_0}\right]_{h_0=0}
	&= \quad
	4J-1 \,,
	& &\qquad (J_c^{(1)}=0.25) \\
	\left[\frac{\partial m^{imp}_2}{\partial h_0}-\frac{\partial m^{imp}_1}{\partial h_0}\right]_{h_0=0}
	&= \quad
	12J^2-4J \,,
	& &\qquad (J_c^{(2)}=0.33333) \\
	\left[\frac{\partial m^{imp}_3}{\partial h_0}-\frac{\partial m^{imp}_2}{\partial h_0}\right]_{h_0=0}
	&= \quad
	\frac{104}{3}J^3-12J^2 \,,
	& &\qquad (J_c^{(3)}=0.34615) \\
	\left[\frac{\partial m^{imp}_4}{\partial h_0}-\frac{\partial m^{imp}_3}{\partial h_0}\right]_{h_0=0}
	&= \quad
	92J^4-\frac{104}{3}J^3 \,.
	& &\qquad (J_c^{(4)}=0.37681)
\end{alignat}
\label{eq:dmdh}
\end{subequations}
It is seen that as we increase the order, 
the approximate value of the critical point 
moves closer to the exact value, 
$J_c=\frac{1}{2}\ln(1+\sqrt{2})=0.440\cdots$. 

It is interesting that the quantity 
on the left-hand side of (\ref{eq:dmdh}) 
is equal to 
the second derivative of the improved free energy with respect to $h_0$ 
at the origin:
\begin{equation}
	\frac{\partial^2 f^{imp}_i}{\partial h_0^2}\Bigg|_{h_0=0}
	=
	\left[\frac{\partial m^{imp}_i}{\partial h_0}
	-\frac{\partial m^{imp}_{i\!-\!1}}{\partial h_0}\right]_{h_0=0} \,.
\end{equation}
Thus, the phase transition point can be understood in terms of the improved 
free energy in the following way. 
The plateau at the origin in the disordered phase is a local minimum, and 
as we increase $J$, it becomes a local maximum, and new minima appear. 
The critical point is the phase transition point. 

It is also interesting to see what happens when 
we turn on an external magnetic field. 
For this purpose, we substitute $h=0.05$, as an example, 
in the improved functions. 
\begin{figure}
\begin{center}
\subfigure[][]{%
	\includegraphics[scale=0.6]{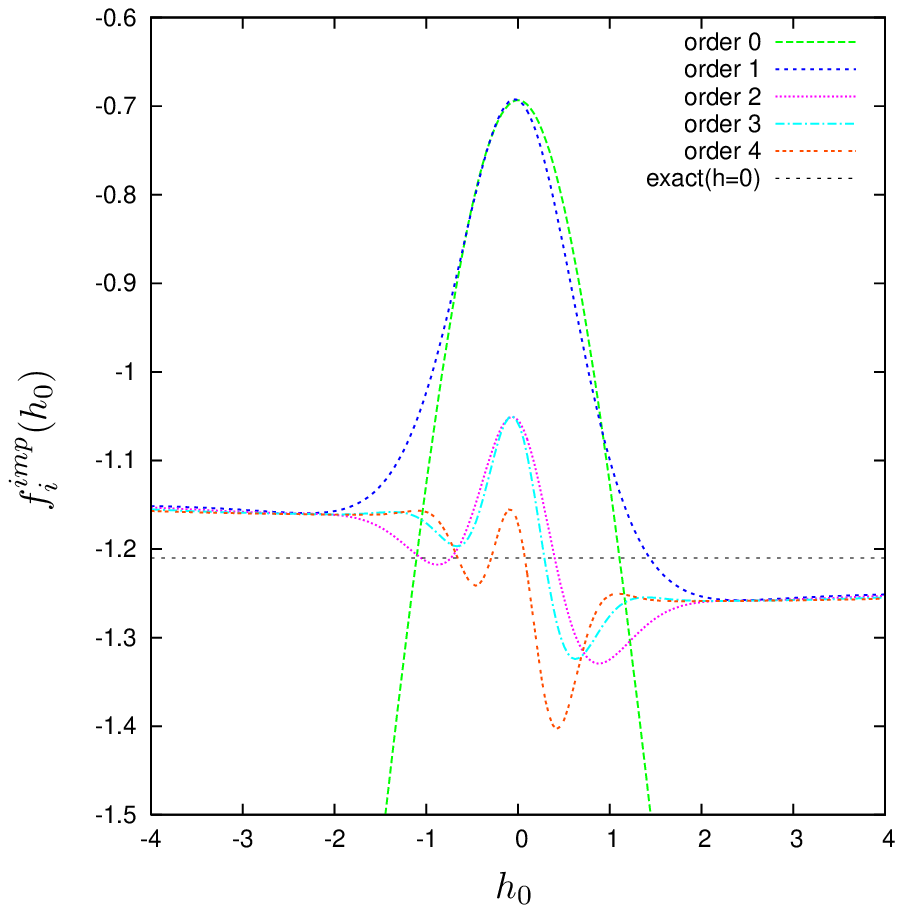}
\label{fig:06}
}
\hskip 2em
\subfigure[][]{%
	\includegraphics[scale=0.6]{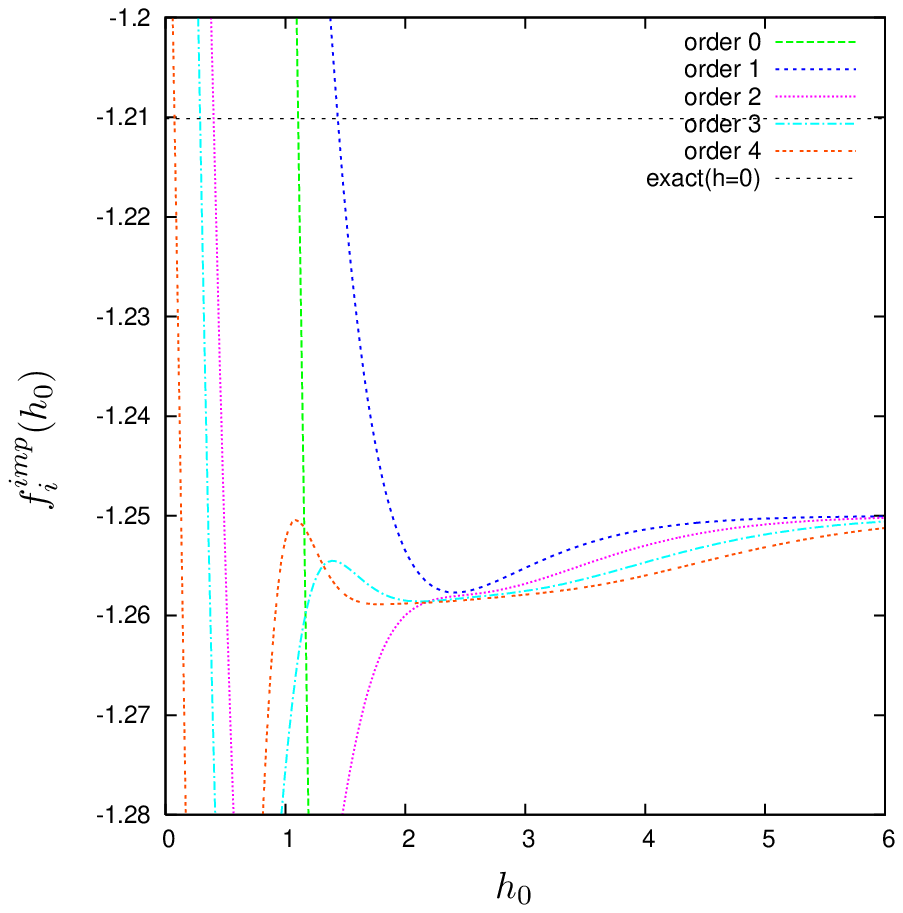}
\label{fig:07}
}
\end{center}
\caption{%
\subref{fig:06} %
Dependences of the improved free energies $f^{imp}_i(h_0)$ 
with respect to the auxiliary parameter $h_0$ 
in the case of a weak external field ($h=0.05$) and $J=0.6$. 
\subref{fig:07} %
An enlarged plot. 
The horizontal line represents the exact value in the case $h=0$. 
}
\label{fig:0607}
\end{figure}
Turning on the external field 
causes the heights of two degenerate plateaux to become different 
[Fig.~\ref{fig:06}]. 
The plateau whose height increases corresponds to 
a metastable physical state (vacuum), 
and the other plateau corresponds to a stable physical state. 
Hence, the ITE reveals not only a stable vacuum but also a metastable one. 
However, as we increase the strength of the external field, 
the metastable plateau disappears, while the stable plateau 
retains its flat shape. 

\vfill

\paragraph{Strong field expansion.}

Let us now proceed to the strong field expansion\footnote{%
The strong field
expansion is closely related to the strong coupling expansion.
Actually, in the strong coupling expansion, 
a small external field is introduced to split the degeneracy 
and to decouple one of the sectors in the thermodynamic limit.
The remaining sector is essentially that obtained from 
the strong field expansion. \cite{Parisi:1988nd}
} 
and construct a series in terms of $H=\exp(-2h)$: 
\begin{equation}
	Z(J,h)\equiv\sum_{n=0}^\infty H^n z_n(h) \,.
\end{equation}

We calculate $z_n$ up to fourth order in $H$:
\begin{subequations}
\begin{align}
	z_0(h)
	&= 
	e^{V(2J+h)} , \\
	z_1(h)
	&= 
	z_0V\lambda^2 , \\
	z_2(h)
	&= 
	z_0V\left[2\lambda^3 +\frac{1}{2}(V-5)\lambda^4\right], \\
	z_3(h)
	&= 
	z_0V\left[6\lambda^4 +2(V-8)\lambda^5 +\frac{1}{3!}\left(V^2-15V+62\right)\lambda^6\right] , \\
	z_4(h)
	&= 
	z_0V\biggl[\lambda^4+18\lambda^5+(8V-85)\lambda^6+(V^2-21V+96)\lambda^7
\biggr.
\nonumber \\ & \qquad
	\left.
	+\frac{1}{4!}\left(V^3-30V^2+323V+714\right)\lambda^8\right] ,
\end{align}
\end{subequations}
where $\lambda=\exp(-4J)$.
The free energy density of $j$th order is defined as 
\begin{equation}
	f_j(J,h)=-\frac{1}{V}\ln\left(\sum_n H^n z_n(h)\right)\Bigg|_{j} .
\end{equation}
We obtain the free energy of fourth order as 
\begin{align}
	f_4(J,h)
	&=
	-(2J+h)-\lambda^2 H
	-\left(2\lambda^3-\frac{5}{2}\lambda^4\right)H^2
	-\left(6\lambda^4-16\lambda^5+\frac{31}{3}\lambda^6\right)H^3
\nonumber \\
	& \qquad
	-\left(\lambda^4+18\lambda^5-85\lambda^6
	+96\lambda^7-\frac{121}{4}\lambda^8\right)H^4 \,,
\end{align}
and the magnetization as 
\begin{align}
	m_4(J,h)
	&=
	1-2\lambda^2H+4\left(-2\lambda^3+\frac{5}{2}\lambda^4\right)H^2
	+6\left(-6\lambda^4+16\lambda^5-\frac{31}{3}\lambda^6\right)H^3
\nonumber \\
	& \qquad
	-8\left(-\lambda^4-18\lambda^5+85\lambda^6-96\lambda^7+\frac{121}{4}\lambda^8\right)H^4 \,.
\end{align}

\begin{figure}
\begin{center}
\subfigure[][]{%
	\includegraphics[scale=0.6]{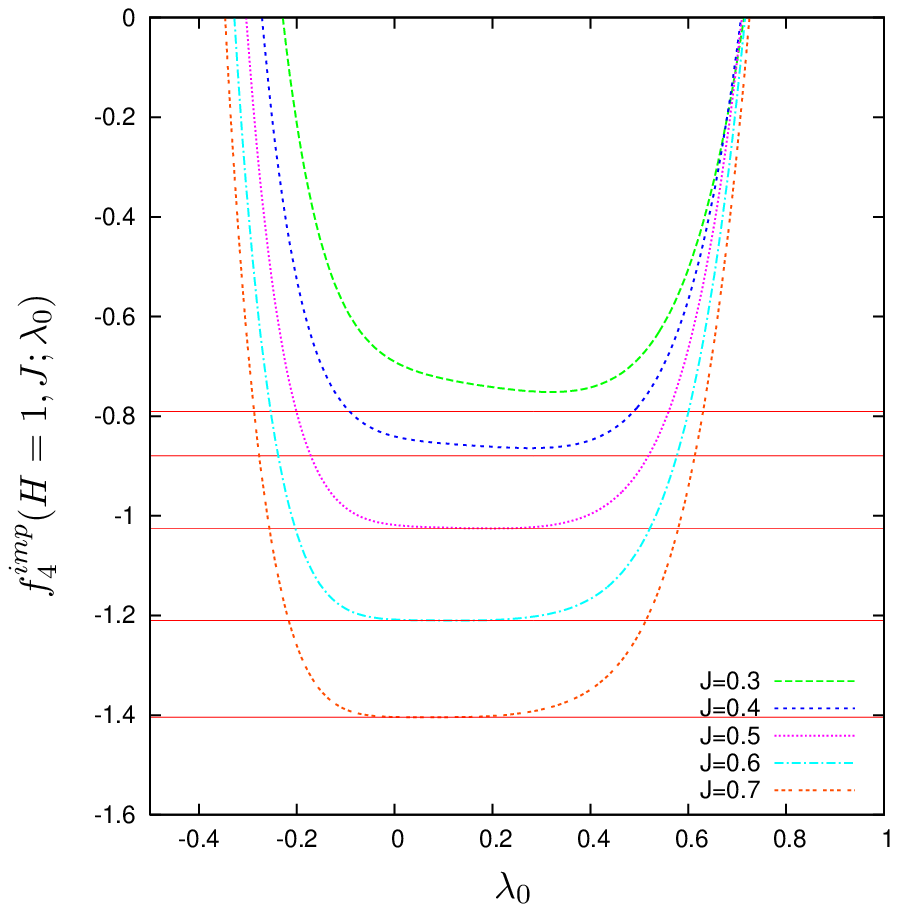}
\label{fig:08}
}
\hskip 2em
\subfigure[][]{%
	\includegraphics[scale=0.6]{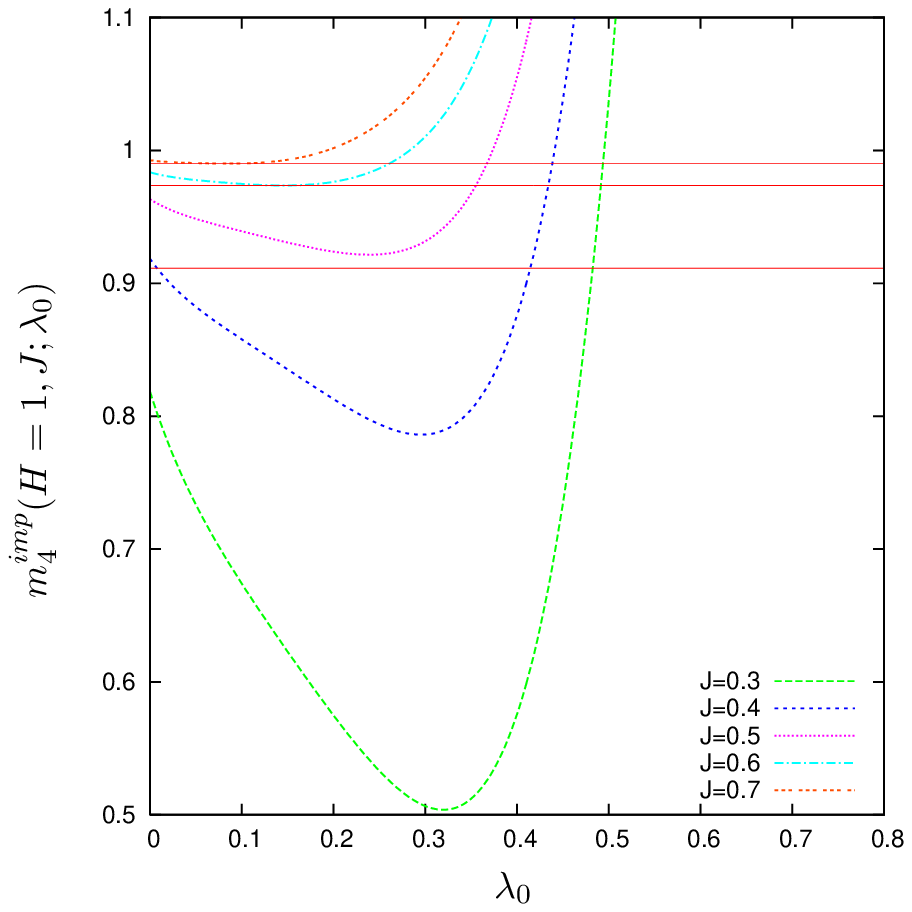}
\label{fig:09}
}
\end{center}
\caption{%
\subref{fig:08} %
Dependences of the fourth-order improved free energies 
$f^{imp}_4(H=1,J;\lambda_0)$. 
\subref{fig:09} %
Those of the improved magnetization $m^{imp}_4(H=1,J;\lambda_0)$ 
in the strong field expansion 
for $J=0.3, 0.4, 0.5, 0.6, \text{and\ } 0.7$.}
\end{figure}

In order to obtain improved functions, 
we make the substitutions 
\begin{align}
	H       & \to gH \,,
\nonumber \\
	\lambda & \to \lambda_0+g(\lambda-\lambda_0)
\end{align}
in the above expressions for the free energy and magnetization. 
Expanding about $g=0$ up to fourth order and setting $g=1$ at the end, 
we obtain the improved free energy $f^{imp}_i(H,J;\lambda_0)$ and 
magnetization $m^{imp}_i(H,J;\lambda_0)$. 
We set $H=1$ because we are interested in the case in which 
the external magnetic field is turned off. 
Here we present the fourth-order improved free energy 
$f^{imp}_4(H=1,J;\lambda_0)$ and magnetization $m^{imp}_4(H=1,J;\lambda_0)$ 
for various $J$ in Figs.~\ref{fig:08} and \ref{fig:09}. 
It is curious that there is no clear transition in the shape of the function 
as we vary the strength of $J$, in contrast to the previous case. 
Nonetheless, an indistinct signal can be seen that the plateau located near 
$\lambda_0 = 0$ moves to $\lambda_0 > 0$ below a certain value of $J$, 
which may imply the occurrence of a phase transition.

\paragraph{Conclusion.}

We applied the improved Taylor expansion (ITE) 
to the Ising model in two dimensions 
and found that it is effective for extracting quantities in 
strong coupling regions from the weak coupling expansion, even though 
these two regions are separated by a phase transition. 
We obtained a good approximation for 
the observable in the ordered phase from a perturbation theory 
formulated in the disordered phase. 
We also determined approximate values of the 
critical point at each order of the approximation by closely 
examining improved functions of the magnetization, 
in particular at the origin of the auxiliary parameter space. 
Furthermore, turning on a small external field 
the ITE reveals the existence of not only a stable 
vacuum but also a metastable one. 
We also applied the ITE to the strong field expansion 
and found that it also works well for evaluating the free energy and 
magnetization in weak coupling regions, 
even though there is only a weak signal indicating the phase transition, 
which can be seen clearly in the weak coupling expansion. 

We note that the ITE at first order is identical to 
the ordinary mean field approximation when we adopt a criterion 
for the identification of the plateau using the stationary points 
of the improved free energy. 
The condition that the derivative of the improved free energy 
with respect to the auxiliary parameter be zero should here 
play the role of the self-consistency condition in the mean 
field approximation. 
Extensions of the mean field approximation have been 
explored, e.g., in the Bethe approximation, 
in such a way that the quantum degrees of freedom of 
neighboring sites are taken into account step by step. 
This approach incorporates the higher order contributions successively, 
while the physical picture is apparent. 
Our ITE scheme also provides a systematic extension in 
a simple manner based on 
the variational method in a purely mathematical sense. 
It turns out, however, that the ITE is different 
from Bethe approximation already at second order, and 
the physical meaning of the extension is not obvious. 
It would be interesting to clarify the role of the plateau 
and its relation to the self-consistency 
condition in the mean field approximation. 
It would also be interesting to evaluate various critical exponents 
within the ITE scheme as a future work.


\paragraph*{Acknowledgements.}
One of the authors (T.~M.) is grateful to T.~Kuroki for discussions 
at an early stage of this work. 
T.~M. also thanks T.~Koretsune for valuable discussions. 
T.~M. and Y.~S. are supported 
by the Special Postdoctoral Researchers Program at RIKEN. 

\appendix

\section{Improved Functions\label{sec:appendix-A}}

Here we list the improved free energy and magnetization 
up to fourth order in the weak coupling expansion. 
In the following, we employ the quantity $k_0\equiv\tanh(h_0)$. 
\begin{subequations}
\begin{align}
	& f^{imp}_0(J,h;h_0) 
	= 
	-\ln(2\cosh(h_0)) , \\
	& f^{imp}_1(J,h;h_0)
	= 
	-\{\ln(2\cosh(h_0))+k_0(h-h_0)\}-2k_0^2 J , \\
	& f^{imp}_2(J,h;h_0)
	= 
	-\left\{\ln(2\cosh(h_0))+k_0(h-h_0)+\frac{1}{2}(1-k_0^2)(h-h_0)^2\right\} 
\nonumber \\ & \qquad
	-\{(4k_0-4k_0^3)(h-h_0)+2k_0^2\} J
	-\{1+6k_0^2-7k_0^4\}J^2 , \\
	& f^{imp}_3(J,h;h_0)
	= 
	-\left\{\ln(2\cosh(h_0))+k_0(h-h_0)+\frac{1}{2}(1-k_0^2)(h-h_0)^2 \right.
\nonumber \\ & \qquad \left.
	-\frac{1}{3}(k_0-k_0^3)(h-h_0)^3\right\} 
	-\{2k_0^2+(4k_0-4k_0^3)(h-h_0)
\nonumber \\ & \qquad
	+(2-8k_0^2+6k_0^4)(h-h_0)^2\}J
	-\{1+6k_0^2-7k_0^4
\nonumber \\ & \qquad
	+(12k_0-40k_0^3+28k_0^5)(h-h_0)\}J^2
	-\left\{\frac{52}{3}k_0^2-56k_0^4+\frac{116}{3}k_0^6\right\}J^3 , \\
	& f^{imp}_4(J,h;h_0)
	= 
	-\left\{\ln(2\cosh(h_0))+k_0(h-h_0)+\frac{1}{2}(1-k_0^2)(h-h_0)^2 \right.
\nonumber \\ & \qquad \left.
	-\frac{1}{3}(k_0-k_0^3)(h-h_0)^3
	+\left(-\frac{1}{12}+\frac{1}{3}k_0^2
        -\frac{1}{4}k_0^4\right)(h-h_0)^4\right\}
\nonumber \\ & \qquad
	-\biggl\{2k_0^2+(4k_0-4k_0^3)(h-h_0)
	+(2-8k_0^2+6k_0^4)(h-h_0)^2 
\nonumber \\ & \qquad \left.
	+\left(-\frac{16}{3}k_0+\frac{40}{3}k_0^3-8k_0^5\right)(h-h_0)^3\right\}J
	-\{1+6k_0^2-7k_0^4
\nonumber \\ & \qquad
	+(12k_0-40k_0^3+28k_0^5)(h-h_0)
	+(6-66k_0^2+130k_0^4-70k_0^6)(h-h_0)^2\}J^2
\nonumber \\ & \qquad
	-\left\{\frac{52}{3}k_0^2-56k_0^4+\frac{116}{3}k_0^6
	+\left(\frac{104}{3}k_0-\frac{776}{3}k_0^3+456k_0^5-232k_0^7\right)
        (h-h_0)\right\}J^3
\nonumber \\ & \qquad
	-\left\{\frac{5}{6}+46k_0^2-\frac{937}{3}k_0^4+526k_0^6
	-\frac{521}{2}k_0^8\right\}J^4 ,
\end{align}
\end{subequations}

\begin{subequations}
\begin{align}
	& m^{imp}_0(J,h;h_0)
	= 
	k_0 , \\
	& m^{imp}_1(J,h;h_0)
	= 
	\{k_0+(1-k_0^2)(h-h_0)\}+\{4k_0-4k_0^3\}J , \\
	& m^{imp}_2(J,h;h_0)
	= 
	\{k_0+(1-k_0^2)(h-h_0)+(-k_0+k_0^3)(h-h_0)^2\}
\nonumber \\ & \qquad
	+\{4k_0-4k_0^3+(4-16k_0^2+12k_0^4)(h-h_0)\}J
	+\{12k_0-40k_0^3+28k_0^5\}J^2 , \\
	& m^{imp}_3(J,h;h_0)
	= 
	\biggl\{k_0+(1-k_0^2)(h-h_0)+(-k_0+k_0^3)(h-h_0)^2 
\nonumber \\ & \qquad \left.
	+\left(-\frac{1}{3}+\frac{4}{3}k_0^2-k_0^4\right)(h-h_0)^3
	\right\}
	+\{4k_0-4k_0^3+(4-16k_0^2+12k_0^4)(h-h_0)
\nonumber \\ & \qquad
	+(-16k_0+40k_0^3-24k_0^5)(h-h_0)^2
	\}J
\nonumber \\ & \qquad
	+\{12k_0-40k_0^3+28k_0^5+(12-132k_0^2+260k_0^4-140k_0^6)(h-h_0)
	\}J^2
\nonumber \\ & \qquad
	+\left\{\frac{104}{3}k_0-\frac{776}{3}k_0^3+456k_0^5-232k_0^7\right\}J^3 , \\
	& m^{imp}_4(J,h;h_0)
	= 
	\biggl\{
	k_0
	+(1-k_0^2)(h-h_0)
	+(-k_0+k_0^3)(h-h_0)^2
\nonumber \\ & \qquad \left.
	+\left(-\frac{1}{3}+\frac{4}{3}k_0^2-k_0^4\right)(h-h_0)^3
	+\left(\frac{2}{3}k_0-\frac{5}{3}k_0^3+k_0^5\right)(h-h_0)^4
	\right\}
\nonumber \\ & \qquad
	+\biggl\{
	4k_0-4k_0^3
	+(4-16k_0^2+12k_0^4)(h-h_0)
	+(-16k_0+40k_0^3-24k_0^5)(h-h_0)^2
\nonumber \\ & \qquad \left.
	+\left(-\frac{16}{3}+\frac{136}{3}k_0^2-80k_0^4+40k_0^6\right)(h-h_0)^3
	\right\}J
\nonumber \\ & \qquad
	+\{
	12k_0-40k_0^3+28k_0^5
	+(12-132k_0^2+260k_0^4-140k_0^6)(h-h_0)
\nonumber \\ & \qquad
	+(-132k_0+652k_0^3-940k_0^5+420k_0^7)(h-h_0)^2
	\}J^2
\nonumber \\ & \qquad
	+\left\{
	\frac{104}{3}k_0-\frac{776}{3}k_0^3+456k_0^5-232k_0^7 \right.
\nonumber \\ & \qquad \left.
	+\left(\frac{104}{3}-\frac{2432}{3}k_0^2+3056k_0^4
	-3904k_0^6+1624k_0^8\right)(h-h_0)
	\right\}J^3
\nonumber \\ & \qquad
	+\left\{
	92k_0-\frac{4024}{3}k_0^3
	+\frac{13216}{3}k_0^5-5240k_0^7+2084k_0^9
	\right\}J^4.
\end{align}
\end{subequations}

\end{document}